\documentclass[aps,pre,twocolumn,showpacs,nofootinbib]{revtex4-1}
\usepackage{enumerate}
\usepackage{amsmath}
\usepackage{amssymb}
\usepackage{graphicx}
\usepackage{epstopdf}
\usepackage{placeins}
\usepackage[T1]{fontenc}

\usepackage{color}

\begin{document}

\title{Efficient fold-change detection based on protein-protein interactions}

\author{W.~Buijsman\textsuperscript{1} and M. Sheinman\textsuperscript{1,2}}

\address{\textsuperscript{1}Department of Physics and Astronomy, VU University, Amsterdam, The Netherlands\\
\textsuperscript{2}Max Planck Institute for Molecular Genetics, 14195 Berlin, Germany}
\date{\today}

\begin{abstract}
Various biological sensory systems exhibit a response to a relative change of the stimulus, often referred to as fold-change detection. In the last few years, fold-change detecting mechanisms, based on transcriptional networks, have been proposed. Here we present fold-change detecting mechanism, based on protein-protein interactions, consisting of two interacting proteins. This mechanism does not consume chemical energy and is not subject to transcriptional and translational noise, in contrast to previously proposed mechanisms. We show by analytical and numerical calculations that the mechanism is robust and can have a fast, precise and efficient response for parameters that are relevant to eukaryotic cells.
\end{abstract}

\maketitle

\section{Introduction}
According to Weber's law, the minimal perceptual change in a stimulus of a sensory system is proportional to the level of that stimulus \cite{Weber1996}. This shows that macroscopic sensory systems, such as vision and hearing, can detect relative changes in stimuli -- a phenomenon referred to as fold-change detection (FCD) \cite{Ferrell2009}. It has been shown that fold-change detection does not apply only to macroscopic sensory systems, but also to certain stimuli in individual living cells employed in multicellular animals, such as \textit{Xenopus laevis} embryos \cite{Goentoro2009a}. At the same time, it was shown theoretically that the incoherent feed-forward loop---a common gene regulation motif appearing often in cells---can provide FCD \cite{Goentoro2009}. Recent experiments find evidences of FCD mechanisms in \textit{E. coli} \cite{Lazova2011, Masson2012}. Recent theoretical work shows that, besides the incoherent feed-forward loop, also the two-state protein can provide FCD \cite{Marquez2011}. Furthermore, it has been shown recently that the incoherent feed-forward loop can provide FCD also in cases with multiple inputs and that FCD is expected to be useful in response to multiple inputs \cite{Hart2013}.

In order to function, FCD mechanisms, based on a transcriptional network, require continues production (transcription and translation) and degradation of transcription factors. As a consequence, this type of mechanisms is subject to transcriptional and translational noises \cite{Raser2005}, continues consumption of chemical energy and lower limit for the FCD response time. The last can be especially significant in mammalian cells ($\sim 20$min) \cite{schroder1989energy,kimura2002transcription,Alon2006}. 

Protein-protein interactions are extensively being studied \cite{xenarios2000dip} and found to play an important role in many aspects of cells' life \cite{schwikowski2000network}, including sensory signal propagation \cite{bren2000signals}. Here, we describe a FCD mechanism, based purely on protein-ligand and protein-protein interactions. Since no gene transcription is involved, it does not consume chemical energy and is not subject to intrinsic transcriptional and translational noises, but only to the variation of the total protein numbers. We analyze the importance of the last quantity in Sec. \ref{sec: examplesrelevance} and show that it is not expected to affect the detection efficiency for eukaryotic cells significantly.  The characteristic response timescales of the mechanism are set by the rates of protein interactions, which have a much broader range and can be much faster than the transcription and translation rates {\cite{Bongrand1999Ligand, Levy2008, Alsallaq2008Electrostatic, Schreiber2009}}. We conclude that FCD, based on protein-protein interactions, can be more effective than the one, based on transcriptional networks. 

This paper contains four sections. Sec. \ref{sec: mechanism} describes the mechanism and shows formally that the mechanism can provide FCD. In Sec. \ref{sec: examplesrelevance} the mean-field action of the mechanism is demonstrated. A discussion of the biological relevance is provided. Sec. \ref{sec: erroranalysis} provides an error analysis of the FCD response and a numerical simulation based on the Gillespie algorithm. In this section, it is shown that the mechanism can be precise, robust and efficient for parameters that are relevant for eukaryotic cells. Sec. \ref{sec: discussionoutlook} provides a discussion of the results.

\begin{figure}[tb]
\centering
\includegraphics[width= \columnwidth]{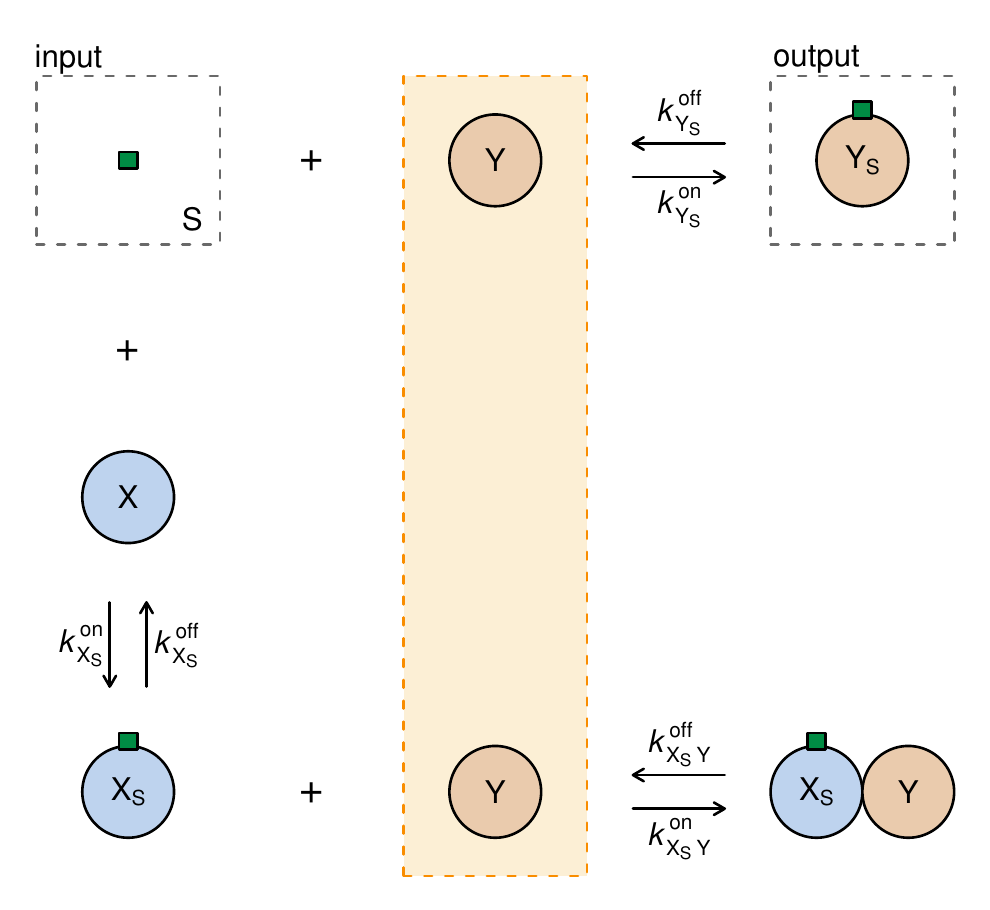}
\caption{A schematic representation of the FCD mechanism, as described in this paper. The cursive $k$-terms indicate the reaction rates, corresponding to the reaction equations, as shown both horizontally and vertically. The ligand $S$ and protein-ligand synthesis product $Y_S$ serve as the input and output, respectively. For clarity, the two copies of $Y$ are grouped by a box.}
\label{fig: 1}
\end{figure}

\section{The mechanism} \label{sec: mechanism} 
The FCD mechanism, described in this paper, is constructed out of two proteins (denoted by $X$ and $Y$) and one ligand (denoted by $S$). Fig. \ref{fig: 1} gives a schematic overview of the reactions and the corresponding reaction rates that are involved. In this Section we, firstly, indicate the parameters requirements for proper FCD. Secondly, it is shown formally that the mechanism can provide FCD. Finally, the response of the mechanism to fold-changes is characterized quantitatively.

\subsection{Mechanism description}
First, consider the reaction $S + X \rightleftarrows X_S$, where $S$ serves as the input signal, and $X_S$ is the complex of $S$ and $X$. Given the association rate $k_{X_S}^\text{on}$ and dissociation rate $k_{X_S}^\text{off}$, the mean-field dynamics is described by
\begin{equation}
\dot{ [X_S]} = [X] [S] \; k_{X_S}^\text{on} - [X_S] \; k_{X_S}^\text{off}. \label{eq: MM1}
\end{equation}
The idea behind the mechanism is to find a parameter regime for which the dynamics of $[S]/[X_S]$ is independent of $[S]_0$ for $[S]$ given by the time-dependent input concentration
\begin{equation}
[S] = \begin{cases} [S]_0 & (t < t_0)\\ \alpha \; [S]_0  & (t \ge t_0) \end{cases}, \label{eq: input}
\end{equation}
where $[S]_0$ is a constant indicating the initial input concentration, $\alpha$ is its fold-change and $t_0$ is the time at which the input concentration fold-changes. Then, an output signal, with a concentration proportional to $[S]/[X_S]$ is generated, by involving the reactions $S + Y \rightleftarrows Y_S$ and $X_S + Y \rightleftarrows X_S Y$. Assume that the total concentration of $Y$ proteins is negligible compared to the concentration of $X_S$ proteins, such that the total concentration of $X$ proteins, called $[X]_0$, can be approximated by $[X]_0 = [X] + [X_S] + [X_S Y] \approx  [X] + [X_S]$. The steady-state of Eq. \eqref{eq: MM1}, solved for $[S] / [X_S]$, is given by
\begin{equation}
\frac{[S]}{[X_S]}\bigg|_{t \to \infty} = \frac{k_{X_S}^\text{off} + [S] \;  k_{X_S}^\text{on} }{[X]_0 \; k_{X_S}^\text{on}}, \label{eq: OS}
\end{equation}
which is independent of $[S]$ for $k^\text{off}_{X_S} \gg [S] \; k^\text{on}_{X_S}$. In this limit, a fold-change given by Eq. \eqref{eq: input} increases both the equilibrium values of $[S]$ and $[X_S]$ by a factor $\alpha$, leaving $[S]/[X_S]$ independent of $[S]_0$. As a result, for a negligible low concentration of $Y$ proteins compared to the concentration of $X_S$ proteins, the dynamics of $[S]/[X_s]$ is independent of $[S]_0$ in the limit $k^\text{off}_{X_S} \gg [S] \; k^\text{on}_{X_S}$. For a summary of the FCD conditions, see the end of this Subsection.

The second step is to design a mechanism providing an output proportional to $[S]/[X_S]$. Consider the reactions $S + Y \rightleftarrows Y_S$ with association rate $k_{Y_S}^\text{on}$ and dissociation rate $k_{Y_S}^\text{off}$, and $X_S + Y \rightleftarrows X_S Y$, with association rate $k_{X_S Y}^\text{on}$ and dissociation rate $k_{X_S Y}^\text{off}$. For the schematics of the reactions involved, see Fig. \ref{fig: 1}. In the limit $[Y]_0 \ll [X_S] $, where $[Y]_0$ denotes the total concentration of $Y$ proteins, the reactions involving $Y$ do not alter the dynamics of Eq. \eqref{eq: MM1}. In this regime, the dynamics of $[Y_S]$ and $[X_S Y]$ are described by
\begin{equation}
\dot{[Y_S]}  \approx [S] [Y] \; k_{Y_S}^\text{on} - [Y_S] \; k_{Y_S}^\text{off}, 
\label{eq: MM2}
\end{equation}
and
\begin{equation}
\dot{ [X_S Y]}  \approx [X_S] [Y] \; k_{X_S Y}^\text{on} - [X_S Y] \; k_{X_S Y}^\text{off}, 
\label{eq: MM3}
\end{equation}
where $[X_S]$ satisfies Eq. \eqref{eq: MM1}. By substituting $[Y]$ from the steady-state of Eq. \eqref{eq: MM3} in the steady-state of Eq. \eqref{eq: MM2}, one gets 
\begin{equation}
\frac{[Y_S]}{[X_S Y]} = \frac{[S] \; k_{Y_S}^\text{on} \; k_{X_S Y}^\text{off}}{[X_S] \; k_{X_S Y}^\text{on} \; k_{Y_S}^\text{off}}, \label{eq: SS1}
\end{equation}
which indicates that $[Y_S] / [X_S Y]$ is proportional to $[S] / [X_S]$. We use $[Y_S]$ as the output of the FCD mechanism. In order to provide FCD, $[Y_S]$ has to be proportional to $[S]/[X_S]$. Under the conditions $[Y] \ll [Y_S] + [X_S Y]$ and $[Y_S] \ll [Y]_0$
(which is equivalent with $[X_S] \; k_{X_S Y}^\text{on} \gg k_{X_S Y}^\text{off}$, $[X_S] \; k_{X_S Y}^\text{on} \gg [S] \; k_{Y_S}^\text{on}$, and $k_{X_S Y}^\text{off} \ll k_{Y_S}^\text{off}$), the lhs of Eq. \eqref{eq: SS1} can be approximated by $[Y_S] / [Y]_0$. As a result $[Y_S]$ is proportional to $[S]/[X_S]$, indicating that the mechanism provides FCD under the conditions mentioned in this Section (summarized below).

In short, the presented mechanism provides FCD if
\begin{enumerate}[a.]
\item $k^\text{off}_{X_S} \gg [S] \; k^\text{on}_{X_S}$, such that $[S]/[X_S]$ remains invariant under fold-changes in $[S]$; \label{cond: a}
\item $[Y]_0 \ll [X_S]$, such that $X + S \rightleftarrows X_S$ is unaffected by the reactions involving $Y$; \label{cond: b}
\item reactions involving $Y$ equilibrate fast compared to $X + S \rightleftarrows X_S$, such that the dynamics of the output are determined totally by $S + X \rightleftarrows X_S$; \label{cond: c}
\item $[X_S] \; k_{X_S Y}^\text{on} \gg k_{X_S Y}^\text{off}$, $[X_S] \; k_{X_S Y}^\text{on} \gg [S] \; k_{Y_S}^\text{on}$ and $k_{X_S Y}^\text{off} \ll k_{Y_S}^\text{off}$, such that $[Y_S] / [X_S Y]$ is proportional to $[S] / [X_S]$. \label{cond: d}
\end{enumerate}

\subsection{Conditions for FCD}
Eqs. (5) and (6) of Ref. \cite{Shoval2010} state a set of conditions sufficient for FCD. The properties of the FCD mechanism as proposed in this paper are checked against these conditions in order to show that the mechanism can provide FCD when conditions (\ref{cond: a}-\ref{cond: d}) are satisfied. A mechanism with input $[S]$, internal variable $[Y_S]$ and output $[X_S]$, described by
\begin{align}
& \dot{[X_S]} = f \big( [X_S],[Y_S],[S] \big), \\
& \dot{[Y_S]} = g \big( [X_S],[Y_S],[S] \big),
\end{align}
provides FCD if
\begin{align}
& f \big( \alpha [X_S],[Y_S], \alpha[S]) = \alpha \; f([X_S],[Y_S],[S] \big), \label{eq: SH1} \\ 
& g \big( \alpha[X_S],[Y_S],\alpha [S]) = g([X_S],[Y_S],[S] \big), \label{eq: SH2}
\end{align}
for any $\alpha>0$. Eq. \eqref{eq: SH1} can be interpreted as the statement that if a fold-change $[S] \to \alpha [S]$ leads to $[X_S] \to \alpha [X_S]$ in equilibrium, the dynamics of $[X_S]$ scale linear with $[S]$. Eq. \eqref{eq: SH2} can be interpreted as the statement that the output dynamics depends only of the ratio of $[X_S]$ and $[S]$, and not on their absolute value. Note that the dynamics of the mechanism can be described in terms of $[S]$, $[Y_S]$ and $[X_S]$. In the FCD regime, $[X]$ is given by $[X]_0 - [X_S]$ by condition \eqref{cond: b}. By condition \eqref{cond: d}, $[Y]  \ll [Y_S], [X_S Y]$ and hence $[X_S Y] = [Y]_0 - [Y_S]$.  From \eqref{eq: MM1} and condition \eqref{cond: a}, it follows that, in the FCD limit,
\begin{equation}
\dot{[X_S]} = [S] [X]_0 \; k^\text{on}_{X_S} - [X_S] k^\text{off}_{X_S},
\end{equation}
which satisfies condition \eqref{eq: SH1}. By substituting the approximation $[Y_S] / [X_S Y] \approx [Y_S] / [Y]_0$ in Eq. \eqref{eq: SS1} and using that reactions involving $Y$ are in a quasi-equilibrium, it follows that
\begin{equation}
\dot{[Y_S]} \sim \frac{d}{dt} \bigg( \frac{[S]}{[X_S]} \bigg),
\end{equation}
which satisfies condition \eqref{eq: SH2}. Since both conditions \eqref{eq: SH1} and \eqref{eq: SH2} are satisfied in the FCD regime (conditions (\ref{cond: a}-\ref{cond: d})), it follows that the mechanism can provide FCD.

\subsection{Mean-field FCD}
What is the response of the mechanism to a fold-change?  Let the input concentration $[S]$ be given by \eqref{eq: input}. By substituting $[Y_S] / [X_S Y] \approx [Y_S] / [Y]_0$  in Eq. \eqref{eq: SS1} and using Eq. \eqref{eq: OS}, one can see that the equilibrium value of $[Y_S]$, denoted by  $\tilde{[Y_S]}$, is given by
\begin{equation}
\tilde{[Y_S]} = \frac{[Y]_0 \; k^\text{off}_{X_S} \; k^\text{on}_{Y_S} \; k^\text{off}_{X_S Y}}{[X]_0 \; k^\text{on}_{X_S} \; k^\text{on}_{X_S Y} \; k^\text{off}_{Y_S}},
\end{equation}
and that the amplitude of the response, $\max ([Y_S] - \tilde{[Y_S]})$, for a fold-change by a factor $\alpha$ is given by $(\alpha - 1) \times \tilde{[Y_S]}$. 

The response is characterized by a raising and a decay timescale. By evaluating $\dot{[X_S]}$ at $t = t_0$, one can show that the decay timescale $\tau_d$ is given by
\begin{equation}
\tau_d = \big[ (\alpha - 1) \; k^\text{off}_{X_S} \big]^{-1}, \label{eq: td}
\end{equation}
which can take values less than a second for biologically relevant parameters, as discussed in the next Section. The raising timescale, by condition \eqref{cond: c}, is much smaller than the decay timescale.

\begin{figure}[tb]
\centering
\centering\includegraphics[width= \columnwidth]{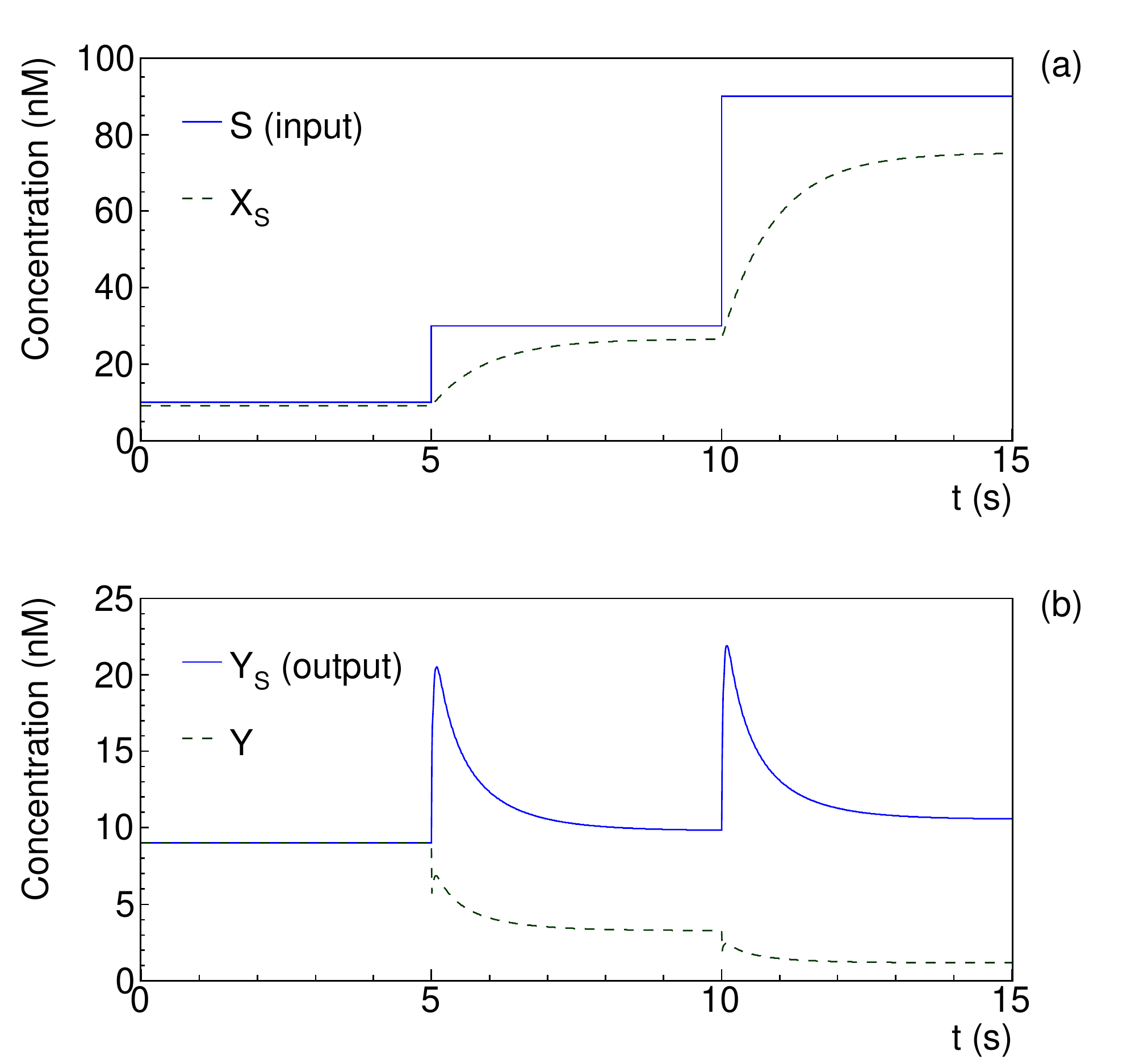}
\caption{Mean-field concentrations $[S]$, $[X_S]$ (a), $[Y]$ and $[Y_S]$ (b) for initial conditions $[S]_0 = 10$nM, $[X]_0 =1 \mu$M, $[Y]_0 = 100$nM, $k^\text{on}_{X_S} = 10^6$M$^{-1}$s$^{-1}$, $k^\text{off}_{X_S} = 1 $s$^{-1}$, $k^\text{on}_{Y_S} = 10^{10} $M$^{-1}$s$^{-1}$, $k^\text{off}_{Y_S} = 100 s^{-1}$, $k^\text{on}_{X_S Y} =10^{10}$M$^{-1}$s$^{-1}$ and $k^\text{off}_{X_S Y} = 10$s$^{-1}$ as a function of time. Concentrations that are not mentioned are zero initially. The system is equilibrated before $t=0$s. The input $S$ fold-changes at $t_0=5$s and $t_0=10$s by a factor $\alpha=3$.}
\label{fig: 2}
\end{figure}

\begin{figure}[tb]
\centering
\centering\includegraphics[width= \columnwidth]{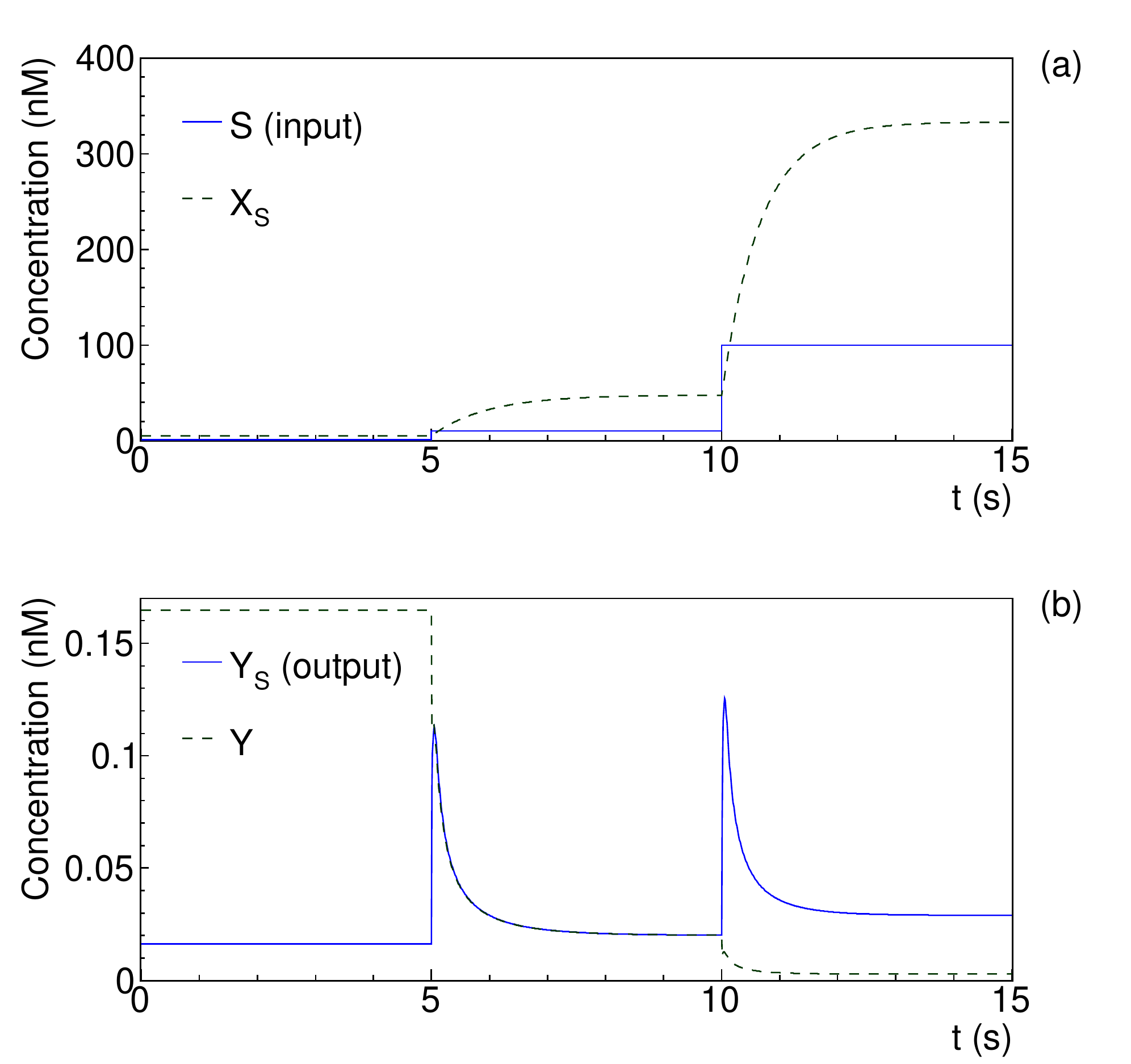}
\caption{A figure identical to Fig. \ref{fig: 2} with parameters $[S]_0 = 1$nM, $[X]_0 =1 \mu$M, $[Y]_0 = 1$nM, $k^\text{on}_{X_S} = 5 \times 10^6$M$^{-1}$s$^{-1}$, $k^\text{off}_{X_S} = 1 $s$^{-1}$, $k^\text{on}_{Y_S} = 10^{10} $M$^{-1}$s$^{-1}$, $k^\text{off}_{Y_S} = 100 s^{-1}$, $k^\text{on}_{X_S Y} =10^{10}$M$^{-1}$s$^{-1}$, $k^\text{off}_{X_S Y} = 10$s$^{-1}$ and $\alpha = 10$.}
\label{fig: 3}
\end{figure}

\section{Examples and biological relevance} \label{sec: examplesrelevance}
It is shown above that the FCD mechanism requires scale separation both for protein-protein and protein-ligand association and dissociation rates and protein concentrations. Affinity constants of protein-protein and ligand-receptor interactions span, over six orders of magnitude, the range $10^{-12}-10^{-6}$M; association and dissociation rates are typically in the $10^3-10^{10}$M$^{-1}$s$^{-1}$ and $10^{-4}-10^{4}$s$^{-1}$ range, respectively \cite{Bongrand1999Ligand,Levy2008,Alsallaq2008Electrostatic,Schreiber2009}. Protein concentrations typically vary in the range $1 - 10^3$nM, corresponding to $10^3 - 10^6$ proteins for eukaryotes and $1 - 10^3$ proteins for prokatyotes per cell \cite{Milo2010}. As shown below, these ranges of biochemical parameters enables fulfilment of the FCD conditions, still leaving a lot of freedom. This freedom, in principle, can be used to tune the timescale of the detection response.

In order to demonstrate the relevance of the mechanism to biological systems, we solve numerically the mean-field equations describing the mechanism dynamics. Fig. \ref{fig: 2} shows an example of mean-field values $[S]$, $[X_S]$, $[Y]$, and $[Y_S]$ as a function of time for an input given by Eq. \eqref{eq: input} with two fold-changes with $\alpha = 3$ at $t_0=5$s and $t_0=10$s. The mechanism parameters are given in the caption; the system is equilibrated before $t=0$s.  In this example, the timescales of the initial response is less than a tenth of a second; the decay timescale is approximately 1s. The figure shows that the mechanism can work for biologically relevant parameters.

A second example is provided in Fig. \ref{fig: 3}. This example, with $\alpha = 10$, shows that the mechanism can work even for high values of $\alpha$. The influence of noise due to the finite number of proteins is discussed in Sec. \ref{sec: erroranalysis}.

\section{Error analysis} \label{sec: erroranalysis}
Figs. \ref{fig: 2} and \ref{fig: 3} show that the response to successive identical fold-changes does not repeat itself perfectly if conditions (\ref{cond: a}-\ref{cond: d}) are satisfied only approximately. This Section first analyzes the response to fold-changes when conditions \eqref{cond: a} and \eqref{cond: d} are not satisfied. Secondly, the robustness of the mechanism is analyzed numerically for the parameters of Fig. \ref{fig: 2}. Finally, a simulation is performed to incorporate the effect of the finite number of proteins in cells in the analysis.

\subsection{Mean-field response when the FCD conditions are not satisfied}
Condition \eqref{cond: a} requires that $[S]/[X_S]$ depends on $[S]_0$ for an input given by Eq. \eqref{eq: input}, after which an output $[Y_S]$, proportional to this quantity, is generated. When \eqref{cond: a} is not satisfied the ratio $[S]/[X_S]$ is different before and after a fold-change, as can be seen in Fig. \ref{fig: 2} (a). In terms of the response amplitude, the error $\epsilon_a$ is given by
\begin{equation}
\epsilon_a = \frac{[S] k^\text{on}_{X_S}}{k^\text{off}_{X_S} + [S] k^\text{on}_{X_S}},
\end{equation}
which vanishes in the FCD limit $k^\text{off}_{X_S} \gg [S] k^\text{on}_{X_S}$. 

For a given error $\epsilon_a$, one can define the parameter $\alpha_\text{max}$, denoting the maximum value of $\alpha$ leaving the change in equilibrium value of $[Y_S]$ within $\epsilon_a$ times the response amplitude. For $n$ successive fold-changes, $\alpha_\text{max}$ is given by
\begin{equation}
\alpha_\text{max} = \left[ \frac{(1 - \epsilon_a) \; [S]  k^\text{on}_{X_S}}{\epsilon_a \; k^\text{off}_{X_S}} \right]^{-1/n}, \label{eq: alphamax}
\end{equation}
which takes a value of $\alpha_\text{max} = 2.3$ for the example of Fig. \ref{fig: 2} at $n=2$ and $\epsilon_a = 0.05$. Eq. \eqref{eq: alphamax} shows that a lower value of the input concentration $[S]$ increases the range of $\alpha$ over which the mechanism provides FCD. However, one should note that $[S]$ can only be decreased up to a limited amount due to condition \eqref{cond: b}. Out of limit \eqref{cond: a}, the expression for the decay timescale $\tau_d$ remains identical.

Condition \eqref{cond: d} requires that the output $[Y_S]$ is proportional to $[S]/[X_S]$. The expansion of the steady-state of Eq. \eqref{eq: MM2} in terms of $[S]/[X_S]$ is given by
\begin{equation}
[Y_S] = \beta_1 \frac{[S]}{[X_S]} + \beta_2 \left[ \frac{[S]}{[X_S]} \right]^2 + \mathcal{O} \left[ \frac{[S]}{[X_S]} \right]^3,
\end{equation}
where
\begin{align}
& \beta_1 = \frac{[Y]_0 \; k^\text{on}_{Y_S} \; k^\text{off}_{X_S Y}}{k^\text{on}_{X_S Y} \; k^\text{off}_{Y_S}}, \\
& \beta_2 = \frac{[Y]_0}{[S]} \left[ \frac{k^\text{off}_{X_S Y}}{k^\text{on}_{X_S Y} \; k^\text{off}_{Y_S}} \right]^2 k^\text{on}_{Y_S} \big( k^\text{off}_{Y_S} + [S] k^\text{on}_{Y_S} \big).
\end{align}
Note that the 0'th order term equals zero, and that the second order term vanishes if condition \eqref{cond: d} is satisfied. In the example, the difference between $[Y_S]$ and its first order expansion is $20 \%$.

\begin{figure}[tb]
\centering
\centering\includegraphics[width= \columnwidth]{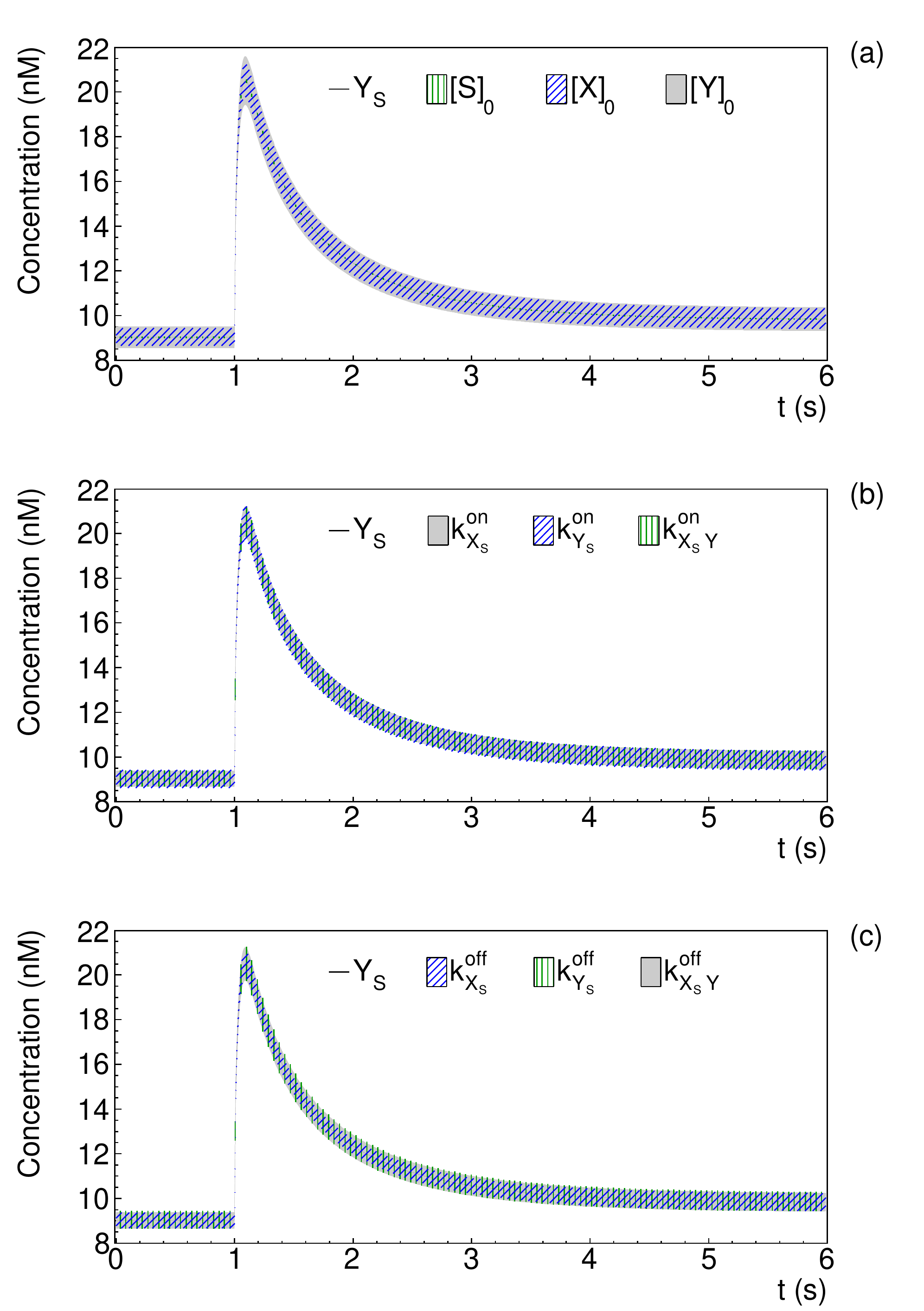}
\caption{The influence of the variation by $\pm 5 \%$ of parameter values on the first response peak of $[Y_S]$ for the parameters of Fig. \ref{fig: 2}. The output range is indicated by a shaded area. The mean-field output $[Y_S]$ is indicated as a black line. In (a), varying $[S]_0$ nearly unaffects $[Y_S]$. Note that most errors differ only slightly from each other. }
\label{fig: 4}
\end{figure}

\subsection{Mechanism robustness}
In this Subsection, we analyze the robustness of the mechanism to parameter variations. Since parameter values, such as reaction rates or concentrations, can depend on external variables such as temperature, it is important to know if the mechanism is robust to parameter variations. The value of each parameter in the example of Fig. \ref{fig: 2} is varied by $5 \%$, after which the mean-field equations describing the mechanism are solved. Fig \ref{fig: 4} indicates the influence of varying parameter values on the first response peak of the output $[Y_S]$. One can see that, in this example, the mechanism is robust to parameter variations.

\begin{figure}[tb]
\centering
\centering\includegraphics[width=\columnwidth]{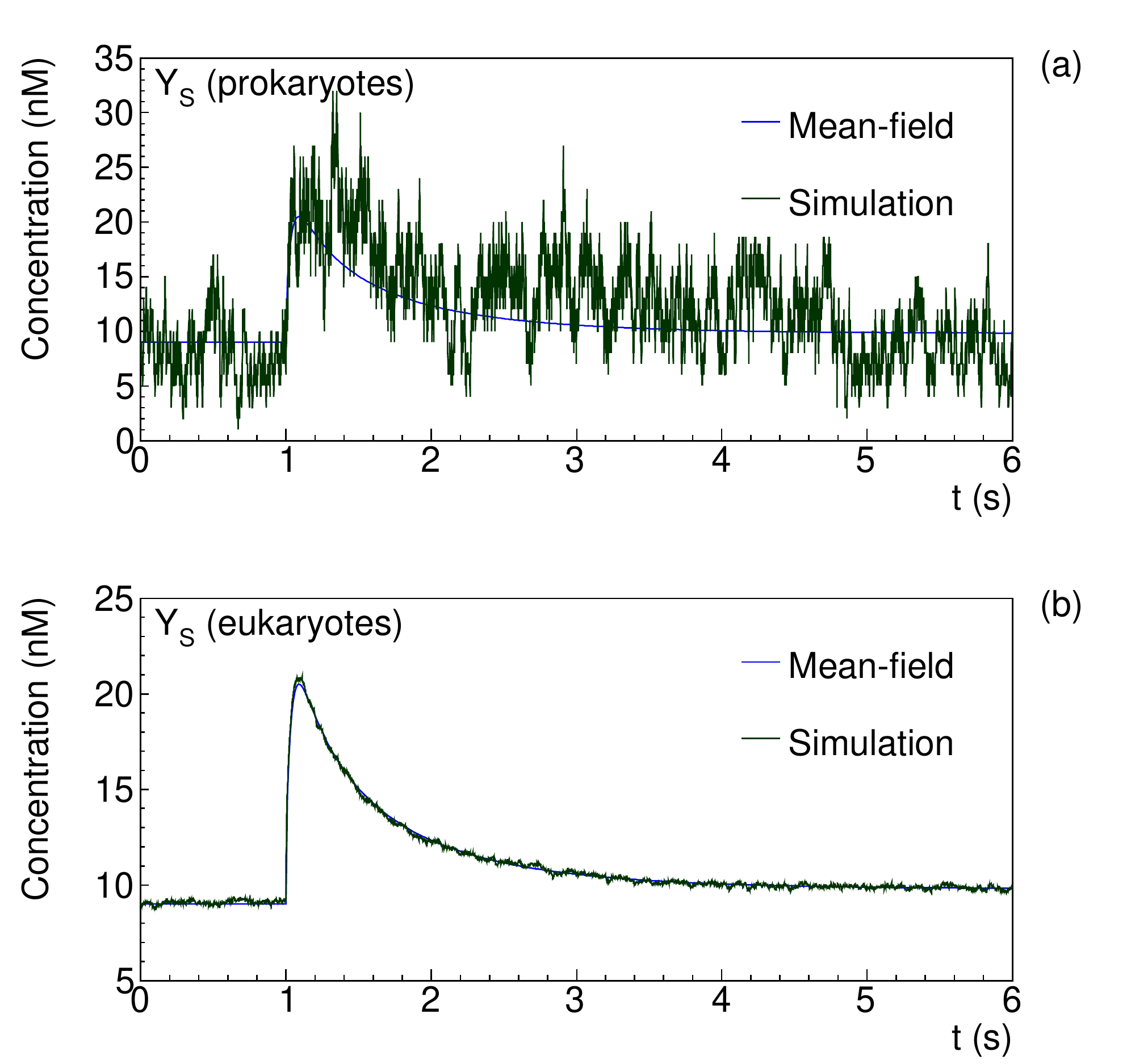}
\caption{Mean-field solution (blue) and simulation (green) results of the output $[Y_S]$ as a function of time for the parameters as mentioned in Fig. \ref{fig: 2}. The simulation results in (a) and (b) correspond to prokaryotes and eukaryotes, respectively. For prokaryotes, 1nM corresponds to $1$ protein per cell. For eukaryotes, where 1nM corresponds to $10^3$ proteins per cell. Note that the simulation does not take external noise into account.}
\label{fig: 5}
\end{figure}

\subsection{Noise analysis}
Due to the finite number of proteins in cells, the mean-field behaviour of concentrations as a function of time does not provide a full description of he mechanism characteristics. In order to investigate the influence of a finite number of proteins, we performed a simulation based on the Gillespie algorithm \cite{Gillespie1977}. Fig. \ref{fig: 5} shows the output concentration $[Y_S]$ as a function of time for the parameters as used in Fig \ref{fig: 2}. The results are presented for protein numbers corresponding to both prokaryotes (a) and eukaryotes (b). For prokaryotes, 1M is roughly equivalent to $10^9$ molecules per cell; for eukaryotes 1M is roughly equivalent to $10^{12}$ molecules per cell. As one can see, for prokaryotic cells the response is too noisy for effective FCD due to finite number statistics. For eukaryotes, the output signal is nearly unaffected by the noise due to large protein numbers. Note that the simulation does not take external noise into account.

\section{Discussion and outlook}
\label{sec: discussionoutlook}
In this paper we describe a new FCD mechanism, based purely on protein-protein and ligand-protein interactions. The mechanism, in contrast to previously proposed mechanisms \cite{Goentoro2009, Marquez2011}, does not require continuous transcription, translation and degradation of proteins. The mechanism avoids transcriptional bursts and other noise sources due to small number bottleneck processes \cite{Raser2005}, and benefits from typically large numbers of proteins in eukaryotic cells. It is shown that the mechanism is robust to parameter variations.

We find that for eukaryotes, the noise due to finite number statistics does not play a significant role. However, for prokaryotes, the typical protein number is not sufficient for a precise FCD on a single cell level. Previously proposed mechanisms are relatively consuming, since the number of proteins has to be large and their lifetime has to be short in order to reduce detection noise and enable a fast response. From this perspective, the mechanism proposed here is advantageous, since it is acting in detailed balance before and well after the transient detection response \cite{Lan2012}. 

We showed that the mechanism is characterized by a response timescale $\tau_d$ given by Eq. \eqref{eq: td}. The value of $\tau_d$, in principle, is limited only by diffusion of proteins \cite{Levy2008}. Thus, as shown in the example, presented in Fig. \ref{fig: 2}, it can be smaller than a second for a biologically relevant set of parameters.

The fold-change detection mechanism proposed in the paper is shown to act in detailed balance. The assumption of detailed balance is used in analytical analysis, where all the concentrations are calculated assuming equilibration of all the reactions in the system and no irreversible reactions are present. Approximate adaptation is achieved using  scale separation of protein concentrations and reaction rates. 

As shown in Ref. \cite{Lan2012}, exact adaptation of a sensory system requires energy consumption. In our case, with no energy consumption, it is evident due to failure of the mechanism beyond a certain window concentration of the stimulus. Consuming chemical energy one can potentially increase the window. Another problem of the presented mechanism is high noise for too small numbers of the proteins, like in prokaryotic cells. This problem also could be potentially resolved by energy consumption, similarly to, say, noise reduction in kinetic proof-reading mechanism \cite{hopfield1974kinetic}. However, in this study we focused on a simple mechanism in detailed balance, since, as we show, even in this simple case the sensory system works well for certain range of parameters:
the window of the stimulus with good adaptation can be made relatively large for relevant biological parameters. Also, the noise level is found to be small relative to the output signal in eukaryotic cells, where the protein numbers is sufficiently large to suppress small numbers noise. 

We expect that the proposed mechanism is only a single example of a large class of biochemical fold-change detectors, based purely on ligand-protein and protein-protein interactions. 

\FloatBarrier
\bibliography{references}

\end{document}